# HIGH TEMPERATURE PHOTOCHEMISTRY IN THE ATMOSPHERE OF HD189733B

**Running Head:** PHOTOCHEMISTRY ON HD189733B
M. R. Line[1], M. C. Liang[2], Y. L. Yung[1]

[1]Division of Geological and Planetary Sciences, California Institute of Technology, Pasadena, CA, 91125, USA
[2]Research Center for Environmental Changes, Academia Sinica, Taipei, Taiwan
[3]Graduate Institute of Astronomy, National Central University, Jhongli, Taiwan
[4]Institute of Astronomy and Astrophysics, Academia Sinica, Taipei, Taiwan

Corresponding Author Email: mrl@gps.caltech.edu

**ABSTRACT**

Recent infrared spectroscopy of hot exoplanets is beginning to reveal their atmospheric composition. Deep with in the planetary atmosphere, the composition is controlled by thermochemical equilibrium. Photochemistry becomes important higher in the atmosphere, at levels above ~1 bar. These two chemistries compete between ~1-10 bars in hot Jupiter-like atmospheres, depending on the strength of the eddy mixing and temperature. HD189733b provides an excellent laboratory in which to study the consequences of chemistry of hot atmospheres. The recent spectra of HD189733b and HD209458b contain signatures of $CH_4$, $CO_2$, CO and $H_2O$. Here we identify the primary chemical pathways that govern the abundances of $CH_4$, $CO_2$, CO and $H_2O$ in the cases of thermochemical equilibrium chemistry, photochemistry, and their combination. Our results suggest that the abundance of these species can be photochemically enhanced above or below the thermochemical equilibrium value, so some caution must be taken when assuming that an atmosphere is in strict thermochemical equilibrium.

# 1. INTRODUCTION

Of the more than four hundred exoplanets discovered thus far, only a small number are transiting hot exoplanets, dubbed "hot Jupiters", from which we can obtain limited spectral information. A variety of chemical species have been detected in hot Jupiter atmospheres. These include atomic species like sodium (Na) (Charbonneau et al. 2002), atomic hydrogen (Vidal-Madjar et al. 2003), atomic carbon and oxygen (Vidal-Madjar et al., 2004), and the molecular species: $CO$, $CO_2$, $H_2O$ and $CH_4$ (Tinetti et al. 2007; Swain et al. 2009a, 2009b). The detection of these species allows us to begin to explore the chemical pathways that control the observed abundances of these species. The species so far identified suggest that hydrocarbon chemistry via $CH_4$ photolysis as well as oxygen and water reactions are important.

The primary chemical pathways that determine chemical abundances in our own solar system are classified as due to either thermoequilibrium chemistry or photochemistry. Ion chemistry may also be important in these hot, highly irradiated atmospheres as it is important in the upper atmospheres of our own solar system planets (Vuitton et al. 2009; Kim & Fox 1994; Friedson et al. 2005). Current atmospheric modeling of hot Jupiter atmospheres typically assume an atmospheric chemical composition consistent with thermochemical equilibrium (Sharp & Burrows 2006; Showman et al. 2009; Burrows et al. 1997; Fortney et al. 2005; Fortney et al. 2010; O'Donovan et al. 2010; Rogers et al. 2009; Marley et al. 2007). Photochemical or other disequilibrium mechanisms, such as quenching, have not received the same attention. (See, however, Cooper & Showman 2006; Zahnle et al. 2009; Liang et al. 2003, 2004). Thermoequilibrium chemistry occurs in high temperature and pressure regimes where chemical timescales are short, typically deep within the atmospheres of giant planets (~1000 bars).

Abundances are determined solely by the thermodynamic properties of compounds in the system via the minimization of the Gibbs free energy (Yung & DeMore 1989). Photochemistry is a disequilibrium process due to UV alteration by the host star. Photochemistry therefore should be important in hot-Jupiter atmospheres, given their proximity to their host stars (Liang et al. 2003).

Liang et al. (2003) were the first to explore the photochemistry that may occur on highly irradiated giant planets through modeling the sources of atomic hydrogen in HD209458b. However, some of the rate coefficients used in that study are unsuitable for these high-temperature regimes, and several key reactions governing the production and loss of $H_2O$ and $CO_2$ were not included. Additionally, better estimates of temperature and vertical mixing profiles can be obtained from more sophisticated GCM simulations.

Zahnle et al. (2009) explored products of sulfur photochemistry and how they may be responsible for the strong UV absorbers that cause thermal inversions as well as the formation of hydrocarbon soot. So far there have been no detections of sulfur species on these hot Jupiters.

The goal of this investigation is to understand the chemistry that produces the observed abundances of $\sim 10^{-4}$, $\sim 10^{-6}$, $\sim 10^{-4}$, and $\sim 10^{-7}$ for CO, $CO_2$, $H_2O$ and $CH_4$, respectively, as detected in the dayside emission spectrum of HD189733b (Swain et al. 2009a) by combining separate photochemical and thermochemical models and then comparing the results to simulations using photochemistry/thermochemistry alone. Furthermore, it has been recently suggested by Madhusudhan & Seager (2009) that there may be as much as 700 ppm of $CO_2$ present in the atmosphere of HD189733b. The discrepancy between this value and the value from Swain et al. (2009a) is due to the assumed vertical distribution of $CO_2$ in the atmosphere (constant, versus high concentration at one pressure level), which is not well constrained. This discrepancy suggests that there is much degeneracy in retrieving temperature and mixing ratio profiles, and that the exact values of the mixing ratios, or their vertical distributions, of the detected species are not well known. In this study, we identify the important mechanisms that govern the abundance of these detected species and their vertical distribution, using HD189733b as an example.

## 2. MODELING

We use both a thermochemical model and a photochemical model to explain the observed abundances of CO, $CO_2$, $H_2O$ and $CH_4$ in the atmosphere of HD189733b. We want to understand the effects that temperature and eddy mixing have on the photochemically derived mixing ratios. We adopt a hot profile representative of dayside temperatures and cool profile representative of night-side temperatures for 30° N from Showman et al. (2009) (Figure 1). We assume isothermal profiles above the upper boundary of the Showman et al. (2009) GCM for the sake of simplicity. These two profiles appear to have a thermal inversion near 1 mbar with a day-night contrast of ~500 K. The use of two T-P profiles will illuminate the day/night contrast of the modeled species. Though HD189733b is not expected to have an inversion, we still choose these T-P profiles because they span the range of hot Jupiter temperature profiles in the literature (Fortney et al. 2006; Tinetti et al. 2007; Burrows et al. 2008), and the existence of an inversion does not significantly affect the major chemical pathways.

In order to determine the thermoequilibrium abundances we use the Chemical Equilibrium with Applications model developed by Gordon & McBride (1994). These abundances at the appropriate lower boundary (explained later) will be used for our lower mixing ratio boundary condition in the photochemical model. Thermochemical calculations require only pressure and temperature along with the relative molar mixing ratios of the atomic species involved in the compounds of interest, in this case C, O and H (no N or S because they have not yet been detected). For the sake of simplicity, and in the absence of any other information, we assume solar abundance of these species ($[C]/[H] \sim 4.4 \times 10^{-4}$, $[O]/[H] \sim 7.4 \times 10^{-4}$, where $[i]$ denotes the concentration of species $i$ (Yung & DeMore 1999 pg. 112). The thermochemical model computes the abundances of all possible compounds formed by those atomic species via a Gibbs Free energy minimization routine (Gordon & McBride 1994). We compute the equilibrium abundances at each pressure-temperature level for our chosen temperature profiles. We would expect to see thermochemical equilibrium abundances in an atmosphere that is not undergoing any dynamical or photochemical processes, or where chemical timescales are much shorter than any disequilibrium timescales (Cooper & Showman 2006; Prinn & Barshay 1977; Smith 1998).

To compute the photochemical abundances of the species of interest, we use the Caltech/JPL-Kinetics 1D photochemical model (Allen et al. 1981; Moses et al. 2005; Gladstone et al. 1996; Yung et al. 1984) for HD189733b. HD189733b is in a 2.2 day period orbiting at 0.03 AU around a K2V star. We use the UV stellar spectrum from HD22049 which is also a K2V star (Segura et al. 2003). The model computes the abundances for 32 species involving H, C and O in 258 reactions including 41 photolysis reactions and includes both molecular and eddy diffusion. The model uses the same hydrocarbon and oxygen chemistry as in Liang et al. (2003)

and Liang et al. (2004) but with high temperature rate coefficients for the key reactions involved in the production and loss of H, $CH_4$, $CO_2$, CO, OH and $H_2O$.  The reaction rates given in the remainder of this paper are taken from Baulch et al. (1992) unless otherwise noted.  We have also added two key reactions involved in the destruction of $H_2O$ and $CO_2$ .  We have not, however, added a complete suite of reactions in order to achieve thermochemical equilibrium kinetically (e.g., Visscher et al. 2010). We do not expect this omission to invalidate our results, as we have included the key chemical pathways that govern the production and loss of the species of interest. The model atmosphere for the photochemical model uses the two temperature profiles described above.  The lower boundary of the photochemical model is important in determining the mixing ratios throughout the atmosphere.  We will estimate this lower boundary using quench level arguments rather than arbitrarily choosing some level.  For more details on quench level estimation we refer the reader to Cooper & Showman (2009), Smith (1998) and Prinn & Barshay (1977).

Eddy and molecular diffusion are key parameters determining the distribution of the abundances in the atmosphere. Eddy diffusion is the primary vertical transport mechanism in our 1D model. The strength of vertical mixing will determine where in the atmosphere the species become chemically quenched, and thus defines the lower boundary conditions for the photochemical model (Prinn & Barshay 1977; Smith 1997).  Following Prinn & Barshay (1977), the transport timescale is given by

$$\tau_{trans} \cong \frac{L^2}{K_z} \quad (1)$$

where $L$ is a vertical length scale typically chosen to be the scale height and $K_z$ is the eddy diffusion coefficient.  The chemical loss timescale of species $i$ is given by

116 $$\tau_{chem,i} = \frac{[i]}{L_i} \quad (2)$$

117 where *[i]* is the concentration of species *i* and $L_i$ is the loss rate of species *i*, typically determined

118 by the bottleneck reaction. The quench level for species *i* is defined where $\tau_{trans} = \tau_{chem,i}$. For

119 levels where $\tau_{trans} < \tau_{chem,i}$ the mixing ratio of species *i* is fixed at the quench level value. For

120 levels below the quench level, the compounds reach chemical equilibrium.

121

122 In order to determine the quench level in the atmosphere HD189733b, we must first estimate the

123 strength of eddy mixing and the timescale for the conversion of CO to $CH_4$ (Griffith & Yelle

124 1999; Prinn & Barshay 1977). The eddy diffusion profile adopted in this model is derived from

125 a globally RMS-averaged vertical wind profile from a GCM (Showman 2010 private

126 communication) and is estimated by

127 $$K_z \sim wL \quad (3)$$

128 where *w* is the RMS-averaged of the vertical wind velocity. Smith (1998) suggests that the

129 appropriate length scale is some fraction of the scale height. Here we assume that it is the scale

130 height, thus giving us an upper limit on eddy diffusion. The GCM derived RMS-averaged

131 vertical winds range from 0 (at ~200 bars) to 7 m/s (~0.8 mbars). The vertical wind is assumed

132 to be constant above this height. Combining this with a typical scale height of ~200 km gives an

133 eddy diffusion of ~$10^{10}$ cm$^2$ s$^{-1}$ (Figure 1). Typical transport timescales from (1) are on the order

134 of ~$10^5$ s.

135

136 The rate-limiting step in the conversion of CO to $CH_4$, and thus the reaction determining the

137 chemical lifetime of CO, is

138 $$H + H_2CO + M \rightarrow CH_3O + M \quad (4)$$

139    (Yung et al. 1988; Griffith & Yelle 1999; Cooper & Showman 2006).  The rate coefficient in

140    reaction 4 has not been measured in the lab, but its reverse reaction rate has been measured to be:

$$k_r = 1.4 \times 10^{-6} T^{-1.2} e^{-7800/T} \text{ cm}^6\text{s}^{-1} \quad (5)$$

142    where $T$ is the temperature at which the reaction takes place (Page et al. 1989).    The forward

143    reaction rate, $k_f$ can be estimated via

$$\frac{k_f}{k_r} = K_{eq} = e^{(G_f - G_r)/RT} \quad (6)$$

145    where $K_{eq}$ is the equilibrium constant for the net thermochemical reaction (Yung et al. 1988)

$$H + H_2CO \leftrightarrow CH_3O \quad (7)$$

147    where $G_f$ and $G_r$ are the Gibbs free energies of the reaction, given respectively by

148    $H[H]+H[H_2CO]-T(S[H]+S[H_2CO])$ and $H[CH_3O]-TS[CH_3O]$ with $H[X]$ being the enthalpy of

149    formation of species X and $S[X]$ being the entropy of species X.  The enthalpies and entropies of

150    the given species can be found in Yung & DeMore (1999) pg 58. With the relevant

151    thermochemical data and equations (5) and (6) we can estimate the forward reaction rate of

152    reaction (4) to b

$$k_f = 5.77 \times 10^{-12} T^{-1.2} e^{3327/T} \quad (8)$$

154    The CO chemical lifetime can then be determined using:

$$\tau_{chem} \sim \frac{[CO]}{k_f [H][H_2CO]} \quad (9)$$

156    where the concentrations of CO, H and $H_2CO$ are determined via the thermochemical model.

157    Upon equating (9) with (1) using the dayside temperature profile we determine the quench level,

158    and thus the lower boundary to be ~3 bars (~1530 K) which is similar to the results of Cooper &

159    Showman (2006) for HD209458b.   This pressure level is much higher than that of Jupiter (~100

160    bars) (Prinn and Barshay 1977) and is similar to that of brown dwarfs (~6 bars) (Griffith & Yelle

161  1999). Choosing a length scale less than the scale height as suggested by Smith et al (1998) can
162  move the quench level to a higher pressure. This is because the chemical timescale in equation 9
163  increases with increasing altitude and lower temperature. Using a length scale of $0.1H$ instead of
164  $H$ moves the quench level to ~8 bars, at where there is very little change in the thermochemical
165  mixing ratios from ~3 bars (Figure 2). Additionally, there is no significant difference in quench
166  level between the nightside and dayside because the two T-P profiles converge near the quench
167  level.
168
169  We assume a zero concentration gradient at the lower boundary in order to allow photochemical
170  products to sink down into the deeper atmosphere except for the observed species of CO, $H_2O$,
171  $CH_4$, $CO_2$,. For these species we fix the mixing ratios to be the thermochemically-derived values
172  at the ~3 bar quench level: $8.41 \times 10^{-4}$, $6.36 \times 10^{-4}$, $4.09 \times 10^{-5}$, and $1.96 \times 10^{-7}$, respectively, for the
173  dayside and $8.39 \times 10^{-4}$, $6.38 \times 10^{-4}$, $4.25 \times 10^{-5}$, and $1.98 \times 10^{-7}$, respectively, for the nightside. We
174  assume a zero flux boundary condition for the top of the atmosphere e.g, little or no atmospheric
175  escape, though this assumption may not be entirely true for atomic hydrogen (Vidal-Madjar et al.
176  2003). This assumption has a negligible effect on the results.
177
178  **3. RESULTS**
179  **3.1 Thermochemical Results**
180
181  The thermochemically derived mixing ratios (relative to $H_2$) are shown in Figure 2. Again, these
182  are the expected mixing ratios if there were no dynamical or photochemical process occurring in
183  the atmosphere, which we know not to be true. If we focus first on the dayside profiles, we can
    see that CO is the dominant carbon bearing species and remains relatively constant with altitude

as do $H_2O$ and $CO_2$. We also notice that $CH_4$ falls off rapidly with increasing altitude (decreasing pressure). We can understand this result by noting that CO, $CH_4$ and $H_2$ abundances are related through the net thermochemical reactions

$$CH_4 + H_2O \leftrightarrow CO + 3H_2 \qquad (10)$$

$$CO + H_2O \leftrightarrow CO_2 + H_2 \qquad (11)$$

Then by Le Chatelier's principle, as the total partial pressure of the atmosphere decreases, the system will want to resist that decrease in order to maintain equilibrium by producing more molecules (smaller molecules), which in this case results in the production of CO and $H_2$. Upon comparing the dayside profiles to the cooler nightside profile, we notice that $CH_4$ becomes more abundant. $CH_4$ is more energetically favorable at lower temperatures and is much more sensitive to the effects of temperature than CO and $CO_2$. We also note that atomic hydrogen is more abundant at warmer temperatures than at cooler temperatures due to the entropy term in the Gibbs free energy. From a thermochemical perspective, we can expect ~10 mbar mixing ratios of the observable species, CO, $H_2O$, $CH_4$ and $CO_2$ to range from: $(2-9) \times 10^{-4}$, $(6-13) \times 10^{-4}$, $(2.6-6758) \times 10^{-7}$, $(4.7-16) \times 10^{-7}$, respectively, due to the day/night contrast. For comparison, the measured dayside emission values from Swain et al (2009a) for CO, $H_2O$, $CH_4$ and $CO_2$ are respectively, $\sim 10^{-4}$, $\sim 10^{-4}$, $\sim 10^{-7}$, and $\sim 10^{-6}$

### 3.2 Photochemical Results

We run four cases of our photochemical model (Figure 3) in order to compare the effects of temperature and photolysis versus no photolysis on the mixing ratios (relative to $H_2$) for H, CO, $H_2O$, $CO_2$ and $CH_4$. In the following subsections we will discuss the important reactions governing the production and loss of each of the relevant species.

### 3.2.1 $H_2O$, OH, and H

The primary reactions that govern the production and loss of $H_2O$ are

R71          $H_2O + h\nu \rightarrow H + OH$          $J_{71}=2.587\times10^{-8}$ s$^{-1}$ (1 mbar)

R137          $OH + H_2 \rightarrow H_2O + H$          $k_{137}=1.70\times10^{-16}T^{1.6} e^{-1660/T}$ cm$^3$ s$^{-1}$

R254          $H + H_2O \rightarrow OH + H_2$          $k_{254}=7.50\times10^{-16} T^{1.6}e^{-9718/T}$ cm$^3$ s$^{-1}$

R137 and R254 are fast enough to readily recycle each other so that the abundance of $H_2O$ remains relatively constant with altitude at the quench level value of $\sim6.36\times10^{-4}$ below the homopause at $\sim10$ nbar. The photolysis of $H_2O$ does not significantly affect its abundance in the observable atmosphere as can be seen in Figure 3, because the loss timescale of $H_2O$ when struck by photolysis is everywhere longer than the transport timescale, thus allowing recently photolyzed parcels to be readily replenished by upwelling. The photolysis of $H_2O$, however, does produce the important OH and H radicals that drive the remainder of the chemistry (Figure 4), with the net result being the conversion of $H_2$ to 2H.

$H_2O$ photodissociates into OH and H at wavelengths lower than 2398 Å. For HD189733b below this wavelength there are $\sim8\times10^{15}$ photons cm$^{-2}$ s$^{-1}$ available for $H_2O$ photolysis. For comparison, the UV flux below this wavelength at Jupiter is $\sim3\times10^{14}$ photons cm$^{-2}$ s$^{-1}$ and for HD209458b, $\sim3\times10^{18}$ photons cm$^{-2}$ s$^{-1}$. OH and H increase with increasing altitude due to the availability of more UV photons. The production of H at high altitudes via $H_2O$ photolysis may be the driver of hydrodynamic escape on hot Jupiters (Liang et al. 2003).

In short, the abundance of $H_2O$ is primarily set by the thermochemical equilibrium value at the lower boundary condition, taken here to be the quench level, and rapidly decreases with altitude above the homopause. If the quench level changes, the observable value of $H_2O$ will change but not significantly, as can be seen in Figure 2. The derived value here is slightly higher than the Swain et al. 2009a dayside emission observations of $(0.1-1) \times 10^{-4}$ but is more consistent with the value obtained by the Tinetti et al. 2007 terminator observations of $\sim 5 \times 10^{-4}$. The day to night contrast is nearly unnoticeable in Figure 3.

### 3.2.2 CO & $CO_2$

Thermochemically, CO is the dominant carbon reservoir in hot atmospheres above $\sim 10$ bars (Figure 2). The abundance of CO is set by the quench level thermochemical equilibrium abundance of $8.4 \times 10^{-4}$. The abundance of $CO_2$ is determined via the interconversion of oxygen from the large reservoirs of CO and $H_2O$ into $CO_2$ via the OH radical. Deeper down in the atmosphere, say, below the quench level, or in the presence of weak vertical transport (low eddy diffusion), oxygen is moved into $CO_2$ via the following reactions

R137 $\quad OH + H_2 \rightarrow H_2O + H \quad k_{137}=1.70 \times 10^{-16} \, T^{1.6} \, e^{-1660/T} \, cm^3 \, s^{-1}$

R152 $\quad OH + CO \rightarrow CO_2 + H \quad k_{152}=1.05 \times 10^{-17} \, T^{1.5} \, e^{250/T} \, cm^3 \, s^{-1}$

R254 $\quad H + H_2O \rightarrow OH + H_2 \quad k_{254}=7.50 \times 10^{-16} \, T^{1.6} e^{-9718/T} \, cm^3 \, s^{-1}$

R255 $\quad H + CO_2 \rightarrow OH + CO \quad k_{255}=2.51 \times 10^{-10} \, e^{-13350/T} \, cm^3 \, s^{-1}$

R152 is the reaction that gives the oxygen from $H_2O$ and CO to $CO_2$. There is no net production or loss of species from these reactions, meaning they will assume thermochemical equilibrium.

Assuming steady state, these 4 reactions can be combined to give the kinetically achieved thermochemical mixing ratio of $CO_2$ in terms of the rate constants ($k$) and mixing ratios ($f$) of the large reservoirs of CO and $H_2O$

$$f_{CO_2} \sim \frac{k_{152}k_{254}}{k_{137}k_{255}} f_{H_2O} f_{CO} \tag{12}$$

$$= 1.85 \times 10^{-7} T^{1.5} e^{5542/T} f_{H_2O} f_{CO}$$

This relation would determine the mixing ratio of $CO_2$ in the absence of any disequilibrium mechanisms such as photochemistry or quenching. Using the thermochemical mixing ratios of $H_2O$ (~$6\times10^{-4}$) and CO (~$9\times10^{-4}$) and evaluating the rate constants at the daytime temperature (T~1200 K) we obtain a $CO_2$ mixing ratio of ~$4\times10^{-7}$ which is consistent with Figure 2.

In the photochemical limit (in the absence of eddy mixing), the photolysis reactions, R71 and R75 become more important and effectively replace R254 and R255, so the important chain of reactions becomes:

R137    $OH + H_2 \rightarrow H_2O + H$    $k_{137}=1.70 \times 10^{-16}\ T^{1.6}\ e^{-1660/T}$ cm$^3$ s$^{-1}$

R152    $OH + CO \rightarrow CO_2 + H$    $k_{152}=1.05 \times 10^{-17}\ T^{1.5}\ e^{250/T}$ cm$^3$ s$^{-1}$

R71     $H_2O + h\nu \rightarrow H + OH$    $J_{71}=2.587\times10^{-8}$ s$^{-1}$ (1 mbar)

R75/76  $CO_2 + h\nu \rightarrow CO + O$    $J_{75/76}=4.4\times10^{-10}$ s$^{-1}$ (1 mbar)

Net     $OH + H_2 \rightarrow 3H + O$

Combining these reactions allows us to estimate the photochemical mixing ratio of $CO_2$ with

$$f_{CO_2} \sim \frac{k_{152}J_{71}}{k_{137}J_{75+76}} f_{H_2O} f_{CO} \tag{13}$$

$$= 0.062 T^{-0.1} e^{1910/T} \frac{J_{71}}{J_{75+76}} f_{H_2O} f_{CO}$$

where $J$ is the photolysis rate of the indicated photolysis reaction. As an extreme case we assume the top of atmosphere photolysis rate of $H_2O$ is $\sim 10^{-5} s^{-1}$, the photolysis rate of $CO_2$ is $\sim 5 \times 10^{-8}$ s$^{-1}$, and the dayside temperature is $\sim$1200 K, giving an upper limit of $\sim$few $\times$ 10$^{-5}$ for $f_{CO2}$. Equation 13 suggests that the abundance of $CO_2$ is photochemically enhanced rather than reduced. The abundance of $CO_2$ in the presence of only quenching (no photochemistry) will remain fairly constant below the homopause at $\sim$1 nbar (Figure 3). This is due to the lack of excess OH produced in R71 used to drive R152 to produce $CO_2$. Again, for comparison, the observed mixing ratio of $CO_2$ from Swain et al. (2009a) is $\sim 10^{-7}$-$10^{-6}$.

### 3.2.3 CH$_4$ and Heavier Hydrocarbons

The primary fate of $CH_4$ in the upper atmosphere is reaction with H to produce $CH_3$, which immediately reacts with $H_2$ to restore $CH_4$,

R28  $\qquad CH_4 + H \rightarrow CH_3 + H_2 \qquad k_{28}=2.20 \times 10^{-20} T^3 e^{-4041/T}$ cm$^3$ s$^{-1}$

R53  $\qquad CH_3 + H_2 \rightarrow CH_4 + H_2 \qquad k_{53}=1.14 \times 10^{-20} T^{2.7} e^{-4739/T}$ cm$^3$ s$^{-1}$

The result is a closed loop. However, the above recycling is not perfect, and the following sequence of reactions occur in the upper atmosphere

R28  $\qquad 2 \times [\, CH_4 + H \rightarrow CH_3 + H_2\,] \quad k_{28}=2.20 \times 10^{-20} T^3 e^{-4041/T}$ cm$^3$ s$^{-1}$

R4  $\qquad 2 \times [\, CH_3 + h\nu \rightarrow CH_2 + H\,] \quad J_{28}=1.95 \times 10^{-3}$ s$^{-1}$ (1 mbar)

R48  $\qquad CH_2 + CH_2 \rightarrow C_2H_2 + 2H \quad k_{48}=1.80 \times 10^{-10} e^{-400/T}$ cm$^3$ s$^{-1}$ (Bauerle et al. 1995)

298    Net            $2CH_4 \rightarrow C_2H_2 + 2H_2 + 2H$

299

300    The net result is production of $C_2H_2$ in the upper atmosphere at the ~1 ppm level. No other $C_2$
301    hydrocarbons are produced in significant quantities. The primary fate of $C_2H_2$ from the upper
302    atmosphere is downward transport, followed by hydrogenation back to $CH_4$.   The abundance of
303    $CH_4$ is ~$4\times10^{-5}$, which is several order of magnitudes larger then the ~$10^{-7}$ detected by Swain et
304    al (2009) and used by Liang et al. (2003).

305

306                                    **4. DISCUSSION**

307    We have analyzed the important disequilibrium mechanisms, photochemistry and simple
308    dynamical quenching that govern the vertical distribution of the observed species in hot Jupiter
309    atmospheres.    With the exception of methane, our derived abundances are consistent with the
310    observations of Swain et al. (2009a). We obtained a value of ~$4\times10^{-5}$, while the observations
311    suggest two orders of magnitude less.    The observed value of ~$10^{-7}$ corresponds to the
312    thermochemical equilibrium value at ~10 mbars.   This would mean the quench level would have
313    to be at this pressure, suggesting an eddy diffusion on the order of ~$10^3$ cm$^2$ s$^{-1}$  from equations
314    (9) and (1).     Alternatively, it may be possible that the observations are probing above the
315    homopause where the mixing ratio can be substantially less than ~$10^{-5}$ (Figure 3).    Line lists
316    used in radiative transfer models are also not well known and are constantly changing at these
317    high temperatures which can play a significant role in dictating the retrieved abundances from
318    the observations (Tinetti 2010 private communication).   Our value of methane is also several
319    orders of magnitude larger than reported by Liang et al. (2003) for HD209458b.  This is because
320    the temperature at the lower boundary used in Liang et al.  (2003) for HD209458b is ~700 K

321  hotter than our lower boundary temperature of ~1530 K and methane is less stable at higher
322  temperatures.
323
324  The metallicity of these hot Jupiters is not well constrained. Swain et al. (2009a) suggests that
325  the metallicty for HD189733b may be subsolar and that the [C]/[O] ratio is between 0.5 and 1.
326  We assumed solar metallicity, but we can explore what might happen if this is not the case.
327  Changes in metallicity will affect the thermochemical equilibrium abundances. This will in turn
328  change the lower boundary mixing ratios. We varied the metallicity (taken here to be
329  ([C]+[O])/[H]) from one tenth solar up to ten times solar to see what effect it would have on our
330  lower boundary mixing ratios (Figure 6). The thermochemical mixing ratios of CO, $H_2O$ and
331  $CO_2$ vary by several orders of magnitude over the range of metallicities, where as $CH_4$ changes
332  very little. This orders of magnitude change at the lower boundary due to metallicity will affect
333  our photochemical results by the same amount. With ten times the solar metallicity we could
334  expect mixing ratios of CO and $H_2O$ to be as high as ~0.1 and $CO_2$ as high as $10^{-5}$. $CO_2$ is more
335  readily affected by metallicity than the other species because it has two oxygen's as opposed to
336  CO's one oxygen. Even higher metallicites will produce more extreme abundances of CO, $CO_2$
337  and $H_2O$.
338
339  The [C]/[O] ratio, also affects the thermochemical abundances. Here we vary the [C]/[O] ratio
340  from 0.1 to 10 times the solar ratio of ~0.6 while keeping the overall metallicity ([C]+[O])/[H])
341  constant at the solar value (Figure 6). The mixing ratio of CO does not vary significantly, but
342  can get as high as ~$10^{-3}$ given a slightly super solar [C]/[O] ratio. $CO_2$ rapidly decreases for
343  ratios above solar and can get as low as 0.1 ppb for 10 times the solar ratio. As the [C]/[O] ratio

344  increases past 1, $H_2O$ and $CH_4$ swap roles in taking up H and can change as much as 3 orders of
345  magnitude.

346

347  There appears to be minor compositional variability between the nightside and dayside.
348  Comparing the solid curves in the top of Figure 3 to the dashed curves in the bottom of Figure 3
349  gives some sense of the magnitude of the day-night variability.  There are no dissociating
350  photons on the nightside, so the quench level mixing and atmospheric circulation determine the
351  abundance throughout the rest of the atmosphere below the homopause.   There is a less than 1%
352  maximum variability in CO and $H_2O$, a factor of ~3 more $CH_4$ on the nightside over the dayside
353  and up to a factor of 2 more $CO_2$ on the dayside.   $CO_2$ and $CH_4$ concentrations experience more
354  variability, because they are most affected by photochemical reactions that only occur on the
355  dayside ($CH_4$ gets destroyed due to R141 and photolysis, $CO_2$ enhanced via equation 13).  $C_2H_2$
356  would exhibit much variability since it is produced strictly from photochemistry.  We could
357  expect to see up to 1 ppm on the dayside of these hot planets with very minute amounts on the
358  nightside where it would be readily thermochemically recycled back to methane.  Terminator
359  observations should fall somewhere between the dayside and nightside values.

360

361                                           **5. CONCLUSIONS**
362  We have shown that both photochemistry and vertical quenching can significantly alter the
363  abundances of $CO_2$, $CH_4$ and $C_2H_2$ in hot Jupiter atmospheres.  Vertical quenching determines
364  the lower boundary values and thus the mixing ratios of CO and $H_2O$, which are not significantly
365  affected photochemically.  $CO_2$ can be photochemically produced above its quench level value
366  by the reaction described in equation (13), and $CH_4$ can be readily photochemically destroyed.

These ideas can be extended to other hot Jupiter atmospheres, though we used HD189733b as our test case. One can see from equation (13) that the fate of $CO_2$ is determined by the temperature of the atmosphere, and the ratio of the $H_2O$ photolysis rate to the $CO_2$ photolysis rate which all depend on the stellar type and the distance. Knowledge of these terms will allow us to predict the abundance of $CO_2$ in any hot Jupiter atmosphere. Finally, the vertical distribution of species derived from thermochemical equilibrium can deviate substantially from those derived via quenching, photochemistry and diffusion, and the simple assumption of thermochemical equilibrium may not be valid in the observable regions of these atmospheres.

**ACKNOWLEDGEMENTS**

We especially thank Adam Showman for providing us his GCM outputs for temperature and vertical winds making it possible for us to determine the eddy diffusion coefficient both of which form the basis of our model atmosphere. We also thank Run-Li Shia, Giovanna Tinetti, Xi Zhang, Konstantin Batygin, Mimi Gerstell, Chris Parkinson, Vijay Natraj, Kuai Le, Mark Swain, Julie Moses, Wes Traub, Pin Chen, Gautam Vasisht, Nicholas Heavens, Heather Knutson, Sara Seager and the Yuk Yung group for very useful discussions and reading the manuscript. MRL was supported by the JPL Graduate Fellowship (JPLGF). MCL was supported in part by an NSC grant 98-2111-M-001-014-MY3 to Academia Sinica. YLY was supported by NASA grant NX09AB72G to the California Institute of Technology.

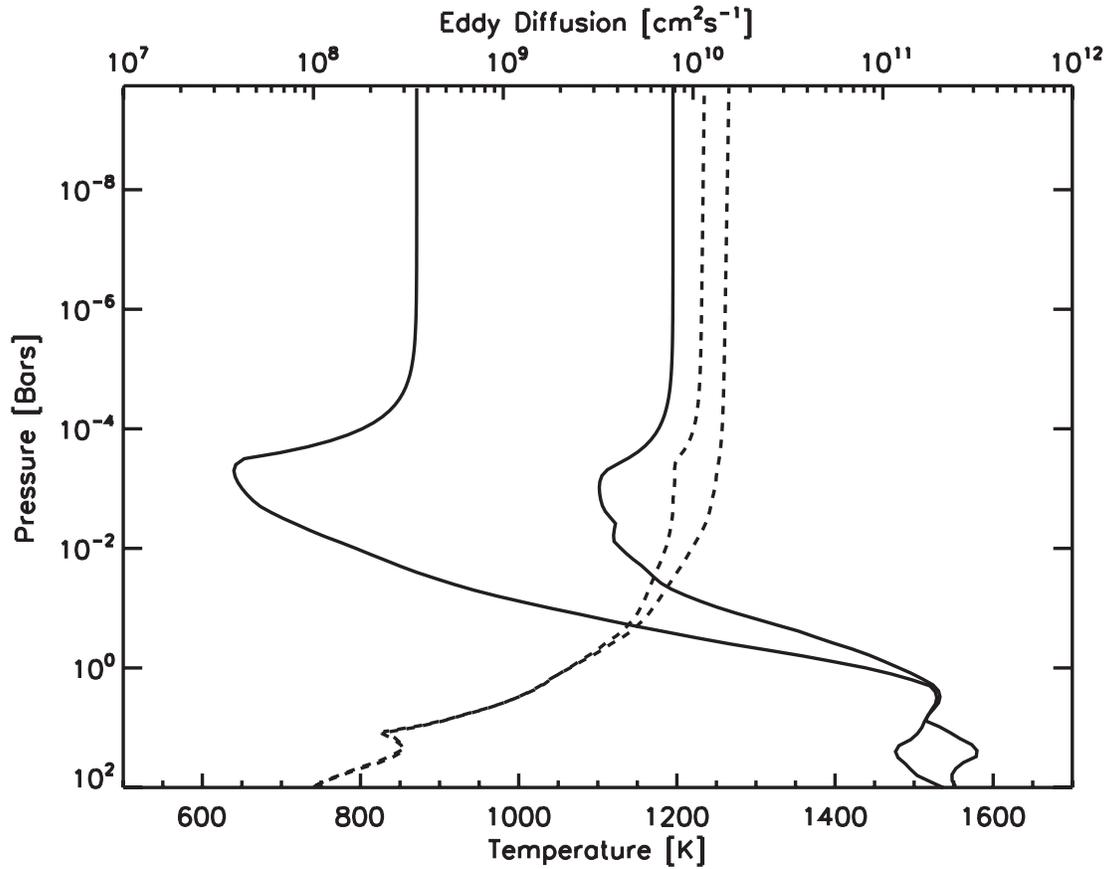

Figure 1. Temperature (solid) and eddy diffusion (dashed) profiles for the model atmosphere. The cooler temperature profile is taken from 30° N from the night side of the model by Showman et al., (2009). The hotter temperature profile is taken from the dayside at the same latitude. The larger eddy diffusion is estimated as discussed in the text (the larger values are for the dayside). Eddy diffusion is read along the top axis, temperature is read along the bottom axis.

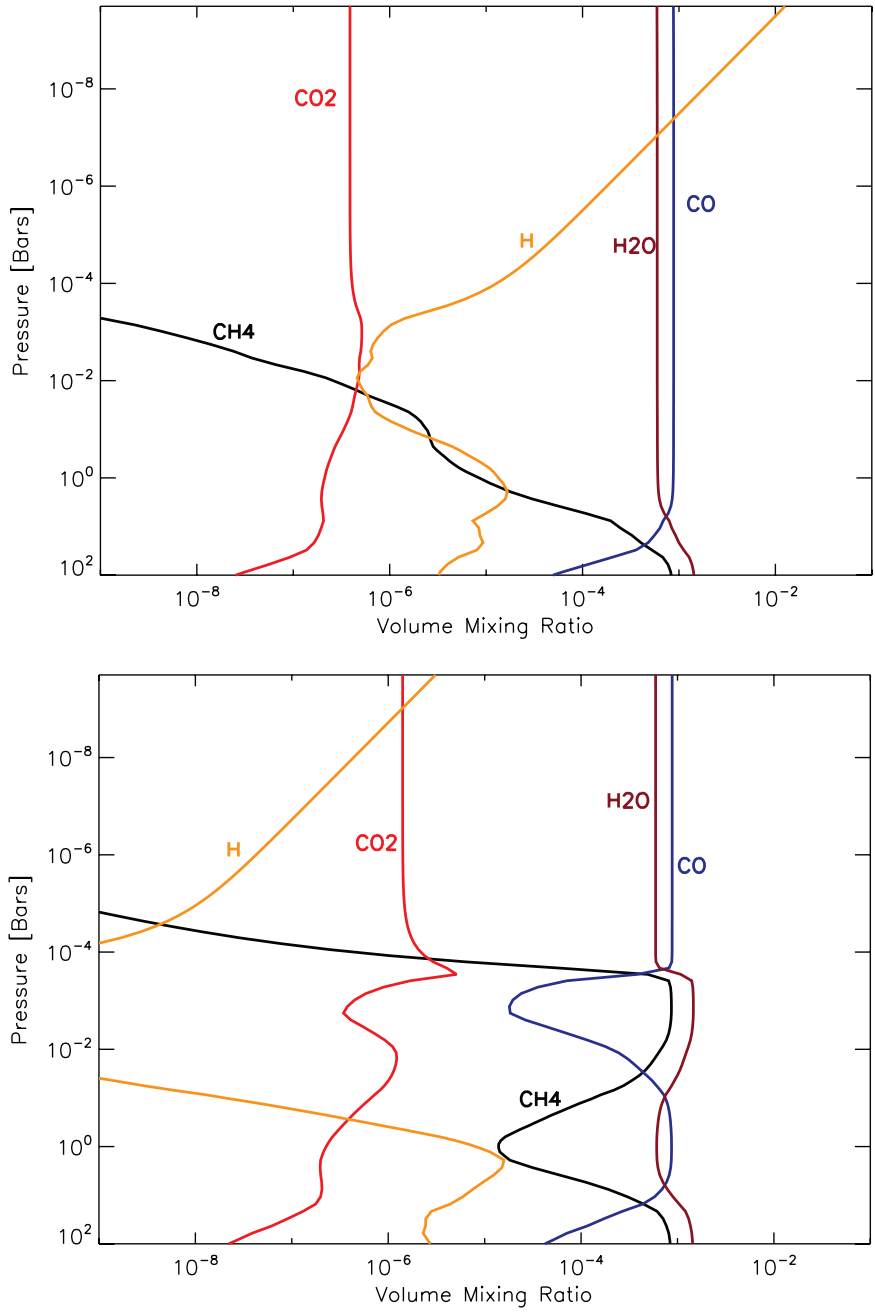

Figure 2. Thermochemical equilibrium mixing ratios derived from the temperature profiles in Figure 1. The top Figure shows the mixing ratios derived for the dayside (hotter) profile. The bottom Figure shows the mixing ratios derived for the (nightside) cooler profile.

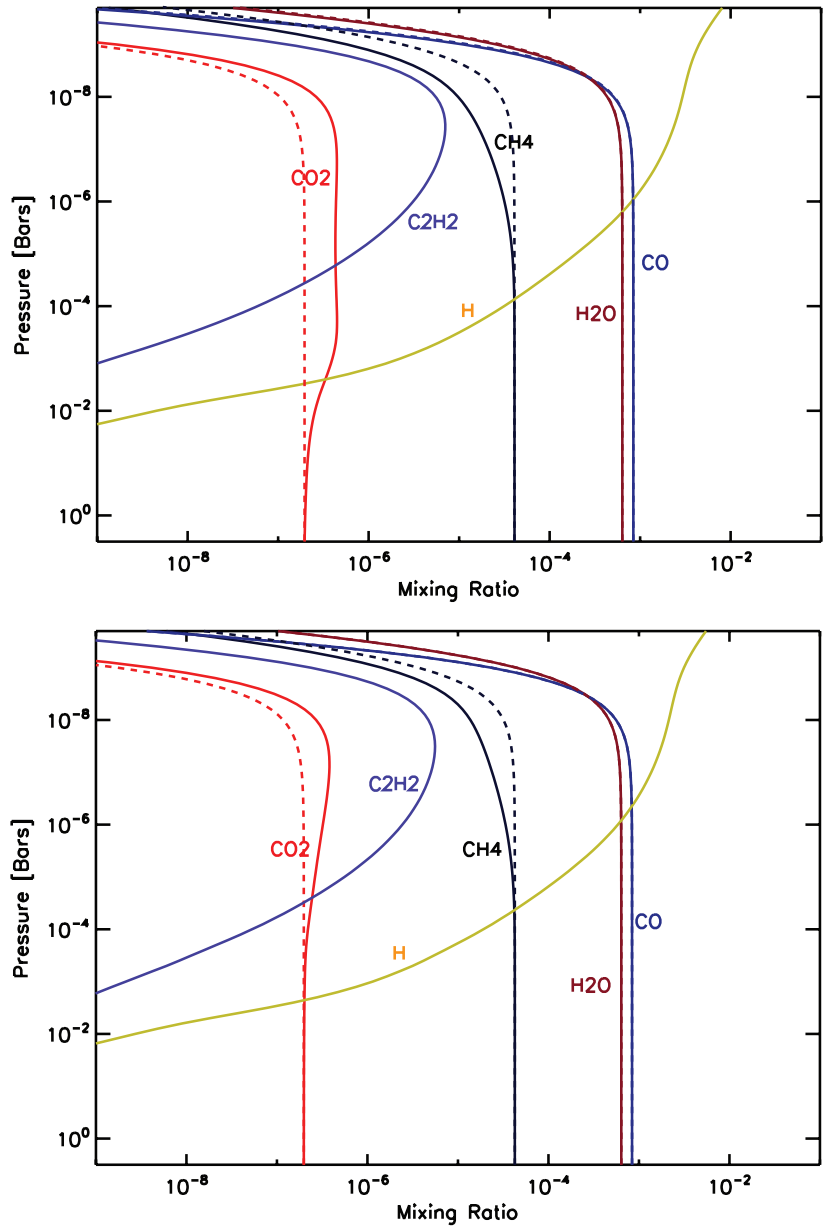

Figure 3. Photochemical mixing ratios (solid) compared to the case with no photochemistry and only quenching (dashed) for the day (top) and night (bottom) temperature profiles. The dashed curves on the bottom plot are representative of what may be seen on the night side of the planet. Note that there is virtually no H or $C_2H_2$ for the cases in which photochemistry is turned off (eg, the dashed curves for these species are not in the plot range).

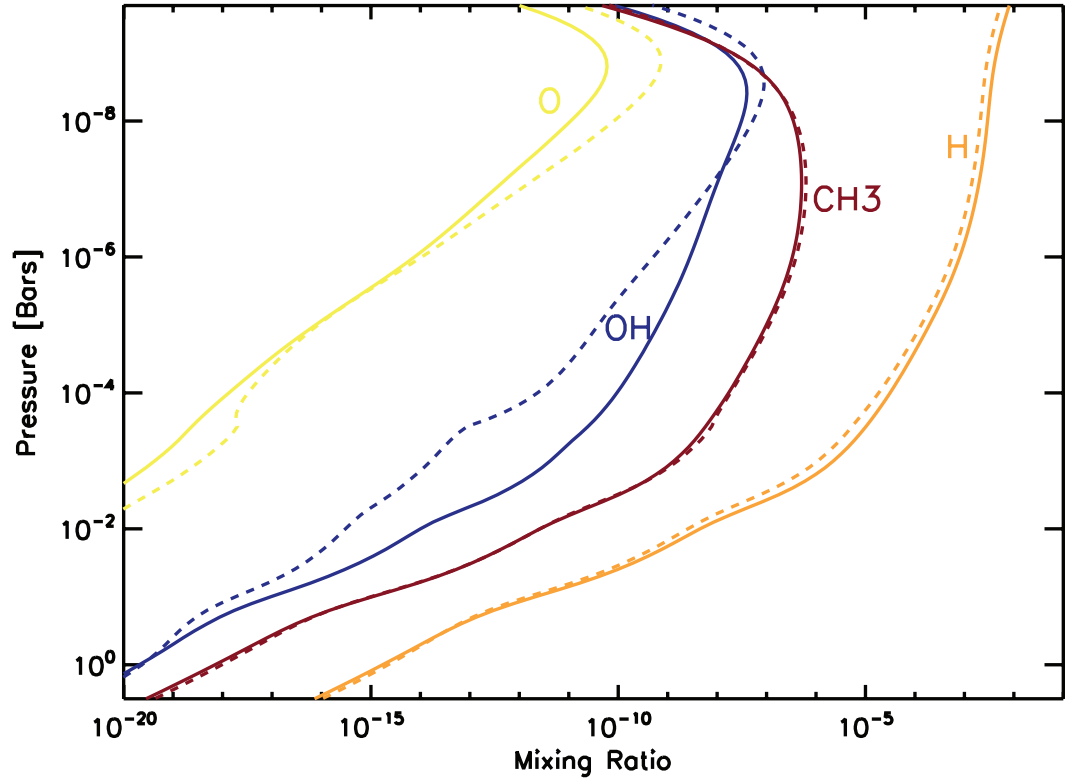

Figure 4. Important radical species involved in pathways governing the abundances of $CH_4$, $H_2O$, CO and $CO_2$. Solid is for the dayside temperature profile, dashed is for the nightside temperature profile. The abundances of radicals increase with decreasing pressure due to the availability of dissociating photons higher in the atmosphere.

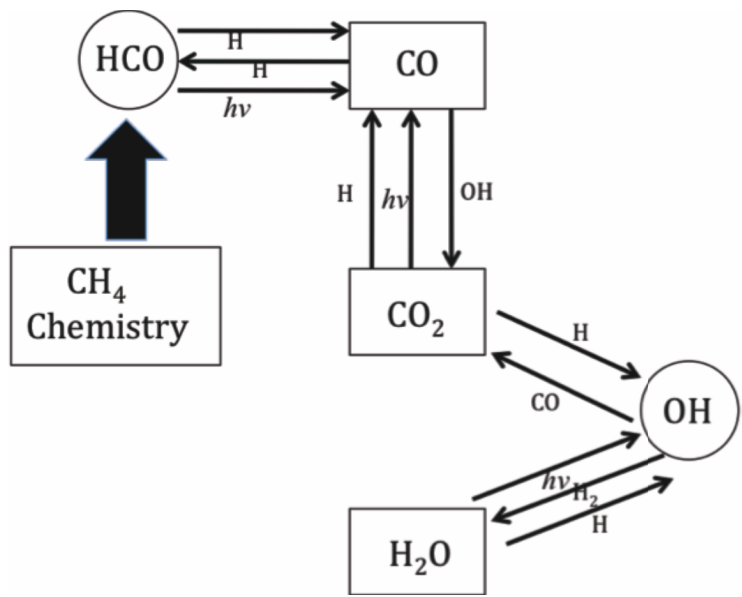

Figure 5. Photochemical web illustrating the important chemical pathways that govern the production and loss of the observable species. The boxes represent the observed species and the circles represent species yet to be observed but are key in the production and loss of the observed constituents.

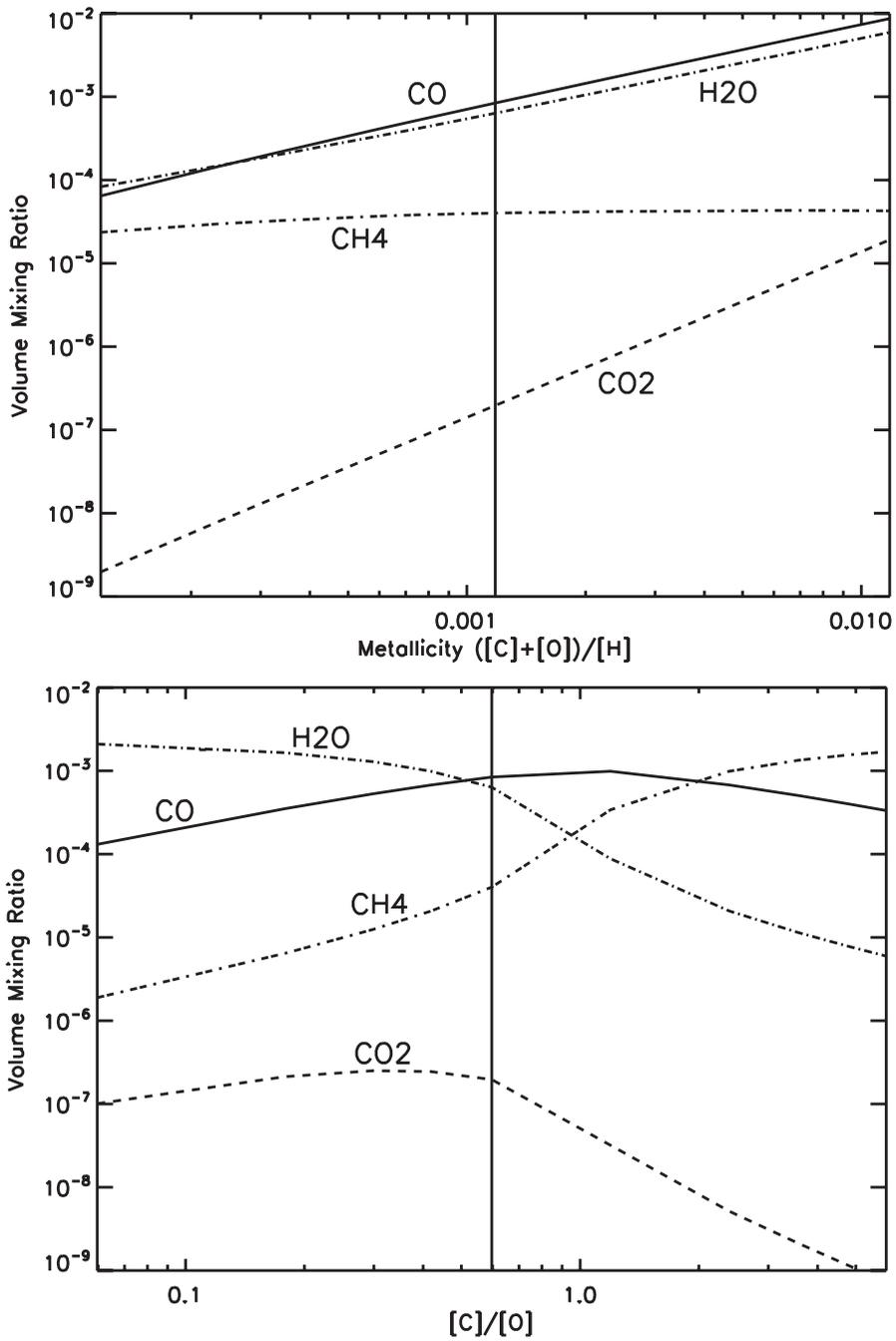

Figure 6. The effects of changing metallicity (top) and C/O ratio (bottom) on the 3 bar quench level mixing ratios for CO, $H_2O$, $CO_2$ and $CH_4$. The vertical lines in each plot represent the solar values.